\documentclass[journal]{IEEEtran}
\usepackage{amsmath,amsfonts}
\usepackage{algorithmic}
\usepackage{algorithm}
\usepackage{array}
\usepackage[caption=false,font=normalsize,labelfont=sf,textfont=sf]{subfig}
\usepackage{textcomp}
\usepackage{stfloats}
\usepackage{url}
\usepackage{verbatim}
\usepackage{graphicx}
\usepackage{cite}
\begin{document}

\title{Pilot-Based Key Distribution and Encryption for Secure Coherent Passive Optical Networks}

\author{Haide Wang, Ji Zhou, Qingxin Lu, Jianrui Zeng, Yongqing Liao, Weiping Liu, Changyuan Yu, and Zhaohui Li
\thanks{Manuscript received; revised. This work was supported in part by the National Natural Science Foundation of China under Grant 62371207 and Grant 62005102 and in part by the Hong Kong Scholars Program under Grant XJ2021018. \it{(Corresponding author: Ji Zhou.)}}

\thanks{Haide Wang, Ji Zhou, Qingxin Lu, Jianrui Zeng, and Weiping Liu are with the Department of Electronic Engineering, College of Information Science and Technology, Jinan University, Guangzhou 510632, China (e-mail: 1834041007@stu2018.jnu.edu.cn; zhouji@jnu.edu.cn; lqx81@stu2022.jnu.edu.cn; jianruizeng@stu2021.jnu.edu.cn; wpl@jnu.edu.cn).}

\thanks{Yongqing Liao is with the China Mobile Internet Co., Ltd, Guangzhou 510640, China (email: 13802885078@139.com).}

\thanks{Changyuan Yu is with the Department of Electronic and Information Engineering, The Hong Kong Polytechnic University, Hong Kong (email: changyuan.yu@polyu.edu.hk).}

\thanks{Zhaohui Li is with the Guangdong Provincial Key Laboratory of Optoelectronic Information Processing Chips and Systems, Sun Yat-sen University, Guangzhou 510275, China, and also with the Southern Marine Science and Engineering Guangdong Laboratory (Zhuhai), Zhuhai 519000, China (e-mail: lzhh88@mail.sysu.edu.cn).}}

\markboth{}%
{Shell \MakeLowercase{\textit{et al.}}: Bare Demo of IEEEtran.cls for IEEE Journals}

\makeatletter
\def\ps@IEEEtitlepagestyle{%
  \def\@oddfoot{\mycopyrightnotice}%
  \def\@evenfoot{}%
}
\def\mycopyrightnotice{%
  {\hfill \footnotesize  \copyright 2023 IEEE\hfill}
}
\makeatother

\maketitle

\begin{abstract}
The security issues of passive optical networks (PONs) have always been a concern due to broadcast transmission. Physical-layer security enhancement for the coherent PON should be as significant as improving transmission performance. In this paper, we propose the advanced encryption standard (AES) algorithm and geometric constellation shaping four-level pulse amplitude modulation (GCS-PAM4) pilot-based key distribution for secure coherent PON. The first bit of the GCS-PAM4 pilot is used for the hardware-efficient carrier phase recovery (CPR), while the second bit is utilized for key distribution without occupying the additional overhead. The key bits are encoded by the polar code to ensure error-free distribution. Frequent key updates are permitted for every codeword to improve the security of coherent PON. The experimental results of the 200-Gbps secure coherent PON using digital subcarrier multiplexing with 16-ary quadrature amplitude modulation show that the GCS-PAM4 pilot-based key distribution could be error-free at upstream transmission without occupying the additional overhead and the eavesdropping would be prevented by AES algorithm at downstream transmission. Moreover, there is almost no performance penalty on the CPR using the GCS-PAM4 pilot compared to the binary phase shift keying pilot.
\end{abstract}

\begin{IEEEkeywords}
Coherent passive optical networks, physical-layer security, advanced encryption standard, pilot-based key distribution, geometric constellation shaping.
\end{IEEEkeywords}

\section{Introduction}
\IEEEPARstart{R}{ecently}, since the standards of 50G passive optical networks (PONs) have been finalized by the International Telecommunication Union, the research on the 100G and beyond PONs has been a hot topic \cite{faruk2021coherent, bonk202250g, zhou2022100g}. Among the solutions, coherent optical technologies with digital subcarrier multiplexing (DSCM) for the beyond 100G PON show competitiveness \cite{zhang2020rate, xu2022intelligent, xing2023first}. Coherent PON using DSCM can combine the time-division multiple access (TDMA) with frequency-division multiple access (FDMA) to provide more end users with higher data rates by the time- and frequency division multiple access (TFDMA). Since physical-layer security issues have always been a concern for the existing PONs at broadcast downstream (DS) transmission, the future coherent PON with higher data rates and more end users will face more severe security issues, especially eavesdropping \cite{kitayama2011security}. Thus, adequate attention should be paid to the physical-layer security enhancement as the transmission performance for coherent PON.

Encryption of cryptography for digital data is the most commonly used means that enhance physical-layer security to combat eavesdropping. Advanced encryption standard (AES) algorithm has been adopted in the standards of the 10G symmetrical PON (XGS-PON) and the 50G-PON \cite{ITUT10Gigabit, ITUThigherspeed}. The data encryption standard and deoxyribonucleic acid algorithms were also demonstrated for physical-layer security enhancement \cite{bi2017chaotic, zhang2018physical, xiao2020novel}. Various physical-layer encryption schemes based on digital chaos have also been investigated \cite{wu2018security, ren2020chaotic, wang2021probabilistic}. However, these new encryption schemes rely on static and pre-shared keys while the key distribution is not solved. Besides encryption, optical steganography based on optical chaos is also used to enhance the physical-layer security of optical communication systems \cite{wang2020scheme}. However, high-speed optical chaos communication faces the challenges of chaos synchronization relying on transceivers with complicated structures, which would be cost-prohibitive for PONs.

Quantum computers pose great challenges to the security enhancement based on classical cryptography, while post-quantum cryptography has made progress in identifying mathematical operations that quantum algorithms cannot significantly speed up \cite{chamola2021information,chawla2023roadmap}. However, post-quantum cryptography faces the challenge of realizing cryptographic usability and flexibility, while maintaining a high level of confidence \cite{bernstein2017post}. Frequent key updates are another way to enhance the security of PON. Even if a key is compromised, the impact is limited to a short period by frequent key updates. Combined with post-quantum cryptography authentication, quantum key distribution can offer extreme security and show promise for future potential \cite{wang2021experimental, mehic2023quantum}. However, quantum key distribution with a relatively low rate is not yet suitable for cost-sensitive PONs with large-scale deployment \cite{vokic2020differential}.

In the previous generation of PONs, keys of the AES algorithm were generated by optical network units (ONUs) and distributed to the optical line terminal (OLT) at the unicast upstream (US) transmission. However, there is at most one frame sublayer header for the key distribution every 125-$\mu$s burst frame. Moreover, the payload and the control messages share the same forward error correction (FEC) codeword, which leads to a decline in the net rate and frequent key updates are not permitted. Therefore, a low-overhead key distribution scheme is required for more frequent key updates to enhance the security of PON. The redundancy of overhead, such as phase and synchronization header was used for key distribution in orthogonal frequency division multiplexing-PON \cite{zhang2021constellation, liang2022secure, li2018secure}. However, FEC coding schemes for error-free redundancy-based key distribution have not been further studied yet. In this paper, we propose the AES algorithm and the geometric constellation shaping four-level pulse amplitude modulation (GCS-PAM4) pilot-based key distribution for the secure coherent PON using DSCM. The main contributions of this paper are as follows:
\begin{itemize}
\item The GCS-PAM4 pilot is designed for both carrier phase recovery (CPR) and key distribution of the AES algorithm without additional overhead and the penalty on CPR. 
\item Key for the AES algorithm is encoded by a polar code and carried by the GCS-PAM4 pilot, which enables error-free and frequent key updates for every codeword.   
\end{itemize}

The rest of the paper is organized as follows. In Section \ref{Pinciple}, the principle of the secure coherent PON is presented. The experimental setups of the secure coherent PON using DSCM are demonstrated in Section \ref{ES}. The experimental results and discussions are given in Section \ref{RESULTS}. Finally, the paper is concluded in Section \ref{CONCLUSION}.

\begin{figure*}[!t]
	\centering
	\includegraphics[width = \linewidth]{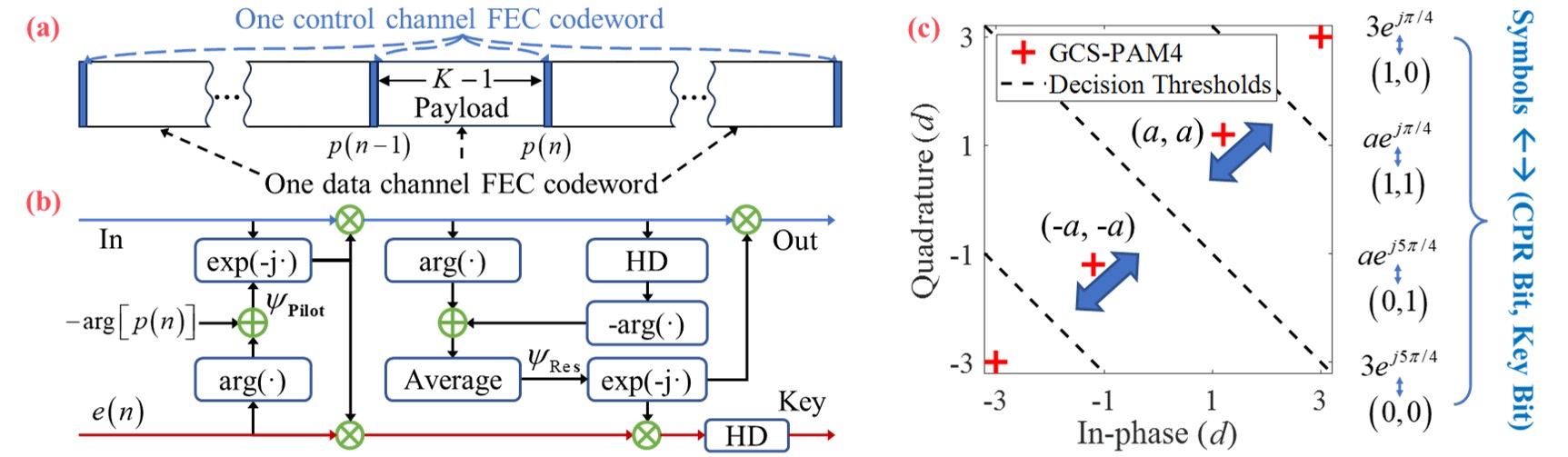}
	\caption{(a) Structure of the data frame with periodically inserted pilot for CPR and key distribution. (b) Diagram of hardware-efficient CPR and key distribution algorithm. (c) The constellation of the GCS-PAM4 pilot with an amplitude coefficient $a$ for CPR and key distribution.}
	\label{fig_GCS}
\end{figure*}

\section{Principle of the secure coherent PON}
\label{Pinciple}
In this section, the principles of the encryption of cryptography, GCS pilot-based CPR and key distribution, and FEC coding for the secure coherent PON are introduced.

\subsection{Encryption of cryptography for secure coherent PON}
Cryptography is the core of security technologies, including encryption, decryption, key distribution, and so on. AES algorithm is one of the most famous symmetric encryption algorithms of cryptography, which was proposed in 2006 \cite{daemen2006understanding}. AES algorithm supports the key of 128, 192, and 256 bits for both encryption and decryption, while the AES-128 and AES-256 have been employed by the XGS-PON and 50G-PON to enhance the physical-layer security, respectively. Therefore, the AES algorithm can be also used in the future coherent PON to enhance the physical-layer security. When a more advanced encryption algorithm is proposed, it can be a candidate encryption scheme for the coherent PON. However, with the advent of quantum computing for brute force attacks, the key for encryption and decryption should be updated more frequently in the coherent PON. Thus, frequent key updates should be achieved without occupying additional overhead.

\subsection{GCS-PAM4 pilot-based CPR and key distribution}

Since the digital signal processing (DSP) for coherent PON should be hardware-efficient, the pilot-based CPR is proposed to replace the blind CPR \cite{zhang2012improved, wang2023fast, li2023pilot}. In the data frame, pilots $p$ are inserted periodically into every $(K-1)$ payload for CPR as shown in Fig. \ref{fig_GCS}(a). The pilot-based CPR and key distribution algorithm is shown in Fig. \ref{fig_GCS}(b), which estimates the phase noise between the $n$-th and the $(n + 1)$-th pilot as
\begin{equation}
\psi_{\text {Pilot}}(n) = \arg \left[e(n)\right] -\arg \left[p(n)\right]
\end{equation}
where $\boldsymbol{e}$ is the received pilot and $\boldsymbol{p}$ is the transmitted pilot. $\operatorname{arg}(\cdot)$ is the angle of a complex. The residual phase noise is estimated as
\begin{equation}
\psi_{\text {Res}}(j)=\frac{1}{2 Q+1} \sum_{l=j-Q}^{j+Q} \left\{\arg \left[q(l)\right] -\arg\left[\hat{q}(l)\right]\right\}
\end{equation}
where $q(l)$ denotes the equalized signal after the $\psi_{\text {Pilot}}(n)$ phase recovery and $\hat{q}(l)$ is $q(l)$ after hard decision (HD). $Q$ is the half-length of the average window. Only the phase of the pilot is pre-shared and used for CPR. Therefore, each ONU can share the secure keys with OLT at the unicast US transmission, which is carried by the amplitude of the pilot and recovered after CPR and hard decision. 

\begin{figure}[!t]
	\centering
	\includegraphics[width =\linewidth]{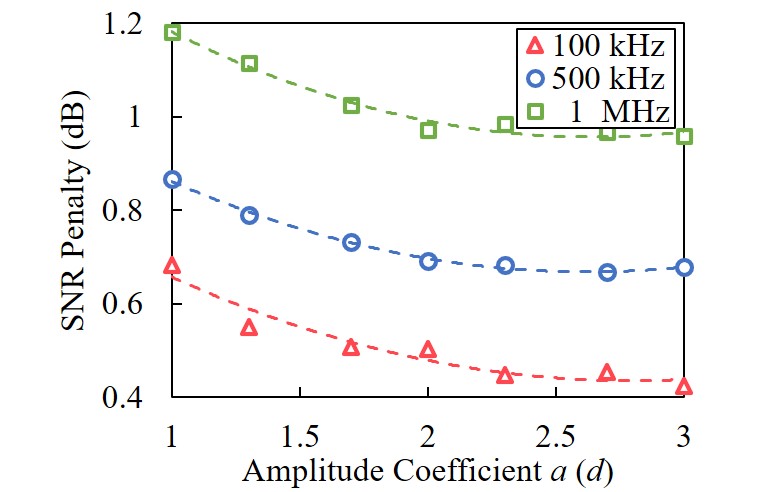}
	\caption{SNR penalty caused by phase noise on 16QAM at the 20\% SD-FEC limit versus the amplitude coefficient $a$ of GCS-PAM4 pilot for CPR.}
	\label{CPR}
\end{figure}

The constellation of the GCS-PAM4 pilot for both CPR and key distribution is shown in Fig. \ref{fig_GCS}(c). The proposed GCS-PAM4 pilot is chosen from the set $\{-3 d\times(1+1j), -ad \times (1+1j), ad \times (1+1j), 3 d\times(1+1j)\}$, where $d$ denotes the unit Euclidean distance and the range of the amplitude coefficient is $0<a\le3$. When $a$ is $1$, the pilot becomes the uniformly distributed PAM4, while it becomes binary phase shift keying (BPSK) if $a$ is $3$. When the Gray coding is used, the de-mapped bits of the GCS-PAM4 pilot are $\{(0, 0), (0, 1), (1, 1), (1, 0)\}$. The first bit represents the phase information for CPR, while the amplitude information represented by the second bit can be used as the control channel for key distribution. The signal-to-noise ratio (SNR) penalty caused by phase noise on 16-ary quadrature amplitude modulation (16QAM) at the 20\% soft-decision (SD) FEC limit (i.e., 2.4$\times 10^{-2}$) versus the amplitude coefficient $a$ of the GCS-PAM4 pilot for CPR is shown in Fig. \ref{CPR}. The linewidth of the laser is set to 100 kHz, 500 kHz, and 1 MHz. As the amplitude coefficient $a$ increases, the phase noise of the laser is estimated more accurately and the SNR penalty is lower. 
\newcounter{TempEqCnt1}
\setcounter{TempEqCnt1}{\value{equation}}
\setcounter{equation}{5}
\begin{figure*}[ht]
	\begin{equation}
	\begin{aligned}
	P_{b, 2} =&\frac{1}{2 \sqrt{\pi N_0}}\left[\int_{k d}^{\infty} e^{-\frac{(x-ad)^2}{N_0}} d x+\int_{-\infty}^{-k d} e^{-\frac{(x-ad)^2}{N_0}} d x +\int_{-k d}^{k d} e^{-\frac{(x-3 d)^2}{N_0}} d x\right]\\
	=&\frac{1}{4}\left[2 \operatorname{erfc}\left(\frac{3-a}{2\sqrt{N_0}}d\right)+\operatorname{erfc}\left(\frac{3+3 a }{2 \sqrt{N_0}}d\right)-\operatorname{erfc}\left(\frac{9+a}{2 \sqrt{N_0}}d\right)\right]
	\end{aligned}\label{secondbit}
	\end{equation}
	\hrulefill
\end{figure*}

The theoretical bit-error ratio (BER) of the first bit in the proposed GCS-PAM4 pilot can be calculated as
\setcounter{equation}{2}
\begin{equation}
\begin{aligned}
P_{b, 1} & =\frac{1}{2 \sqrt{\pi N_0}}\left[\int_0^{\infty} e^{-\frac{(x-a d)^2}{N_0}} d x+\int_0^{\infty} e^{-\frac{(x-3 d)^2}{N_0}} d x\right] \\ &=\frac{1}{4}\left[ \operatorname{erfc}\left(\frac{ad}{\sqrt{N_0}}\right)+\operatorname{erfc}\left(\frac{3d}{\sqrt{N_0}}\right)\right]
\end{aligned}
\end{equation}
where $N_0$ denotes the power spectral density of the noise. The complementary error function is defined as
\begin{equation}
\operatorname{erfc}(x)=\frac{2}{\sqrt{\pi}} \int_{x}^{\infty} e^{-z^{2}} dz.
\end{equation}
$d/\sqrt{N_0}$ can be expressed as
\begin{equation}
\frac{d}{\sqrt{N_0}} = \sqrt{\frac{SNR}{2\times(3^2+a^2)}}.
\end{equation}
The theoretical BER of the second bit in the proposed GCS-PAM4 can be calculated by Eq. (\ref{secondbit}) as shown at the top of this page, where $k = (3+a)/2$ is the normalized decision threshold for the second bit. To achieve a similar CPR performance of the BPSK pilot, the amplitude coefficient $a$ of GCS-PAM4 should be as close to $3$. However, as $a$ increases, the BER of the first bit for CPR becomes smaller, while the BER of the second bit for key distribution becomes larger. Therefore, the amplitude coefficient $a$ should be optimized to trade off the performance of CPR and key distribution.

\begin{figure}[!t]
	\centering
	\includegraphics[width =\linewidth]{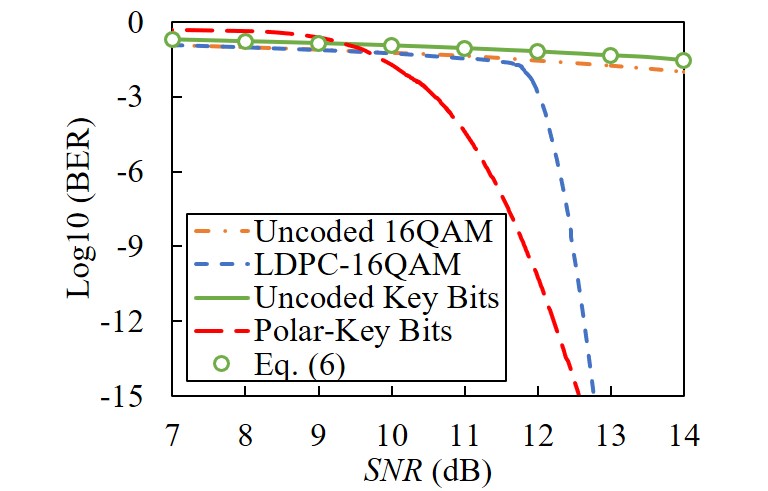}
	\caption{BER versus SNR of the uncoded 16QAM, uncoded key bits, $(17280, 14592)$-LDPC encoded 16QAM, $(512, 256)$-polar encoded key bits, and the theoretical BER of the uncoded key bits given by Eq. (\ref{secondbit}).}
	\label{FEC_perf}
\end{figure}

\subsection{Coding schemes for data and control channels}
Frequent key updates should be permitted for every FEC codeword to improve the security of coherent PON. Therefore, it requires a relatively long FEC codeword for the data channel and a relatively short FEC codeword for the control channel as shown in Fig. \ref{fig_GCS}(a). The $(17280, 14592)$-low-density parity-check (LDPC) code was adopted as the default FEC in 50G-PON \cite{borkowski2021flcs}, which can also be adopted in coherent PON before a more proper coding scheme is proposed. Since the GCS-PAM4 pilot is much shorter than the payload, an FEC code with a short codeword length can be used for the control channel. Referring to the coding schemes standardized by the 3rd generation partnership project (3GPP) for fifth-generation new radio in the enhanced mobile broadband scenarios, the LDPC code can be used for the data channel, while the polar code can be employed for the control channel of coherent PON to realize the error-free key distribution \cite{sharma2017polar, bioglio2020design}. 

Only the second bit of the GCS-PAM4 pilot is used as the control channel to carry the key bit and encoded by polar code. Fig.\ref{FEC_perf} shows the BER versus SNR of the uncoded 16QAM payload, uncoded key bits, $(17280, 14592)$-LDPC encoded 16QAM, $(512, 256)$-polar encoded key bits in simulation and the theoretical BER of the uncoded key bits given by Eq. (\ref{secondbit}). The amplitude coefficient $a$ is set to 1.7 to make a trade-off between the CPR performance and the code rate of the polar code. When the GCS-PAM4 pilot with an $a$ of 1.7 is used for CPR, the SNR penalty is no more than 0.1 dB compared to the BPSK pilot. BER of $(17280, 14592)$-LDPC encoded 16QAM and $(512, 256)$-polar encoded key bits are fitted to reach $10^{-15}$. The key bits in the control channel can reach the BER of $10^{-15}$ in a slightly lower SNR than the data channel.

\section{Experimental setups of secure coherent PON}
\label{ES}
In this section, the experimental setups of the 200-Gbps secure coherent PON using DSCM are introduced. The GCS-PAM4 pilot-based key distribution and the AES algorithm are demonstrated at the US and DS transmissions to enhance the security of coherent PON, respectively.

\begin{figure*}[!t]
	\centering
	\includegraphics[width = \linewidth]{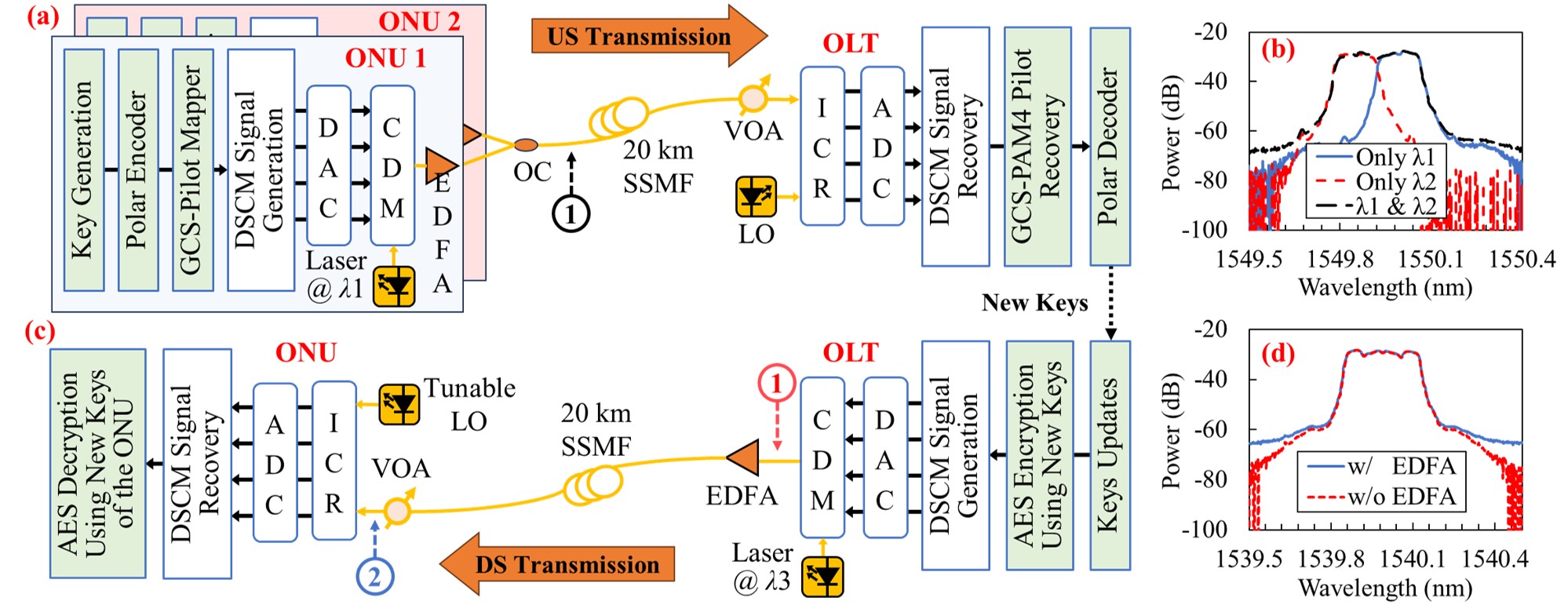}
	\caption{(a) Experimental setups of the secure coherent PON using 4-SCs$\times$8-Gbaud/SC DSCM with the GCS-PAM4 pilot-based key distribution at the unicast US transmission. (b) The optical spectrum of CDM output signal and the US signal at Point 1. (c) Experimental setups of the secure coherent PON using 4-SCs$\times$8-Gbaud/SC DSCM with AES-256 algorithm at the broadcast DS transmission. (d) The optical spectrum of the DS signal at Point 1 and Point 2.}
	\label{ex}
\end{figure*}

\subsection{GCS pilot-based key distribution at US transmission}
The experimental setups of the 200-Gbps secure coherent PON using 4-subcarriers (SCs)$\times$8-Gbaud/SC DSCM with GCS pilot-based key distribution at the unicast US transmission are shown in Fig. \ref{ex}(a). There were two ONUs and each ONU was allocated two subcarriers. The keys for the AES algorithm of each ONU were generated and encoded by the $(512, 256)$-polar code, which uses an 11-bit cyclic redundancy check (CRC) encoding. The encoded key bits were mapped into the GCS-PAM4 pilot as the second bits, while the first bits for the CPR were pre-shared. Then the DSCM signal was generated as follows. The bit sequences of each ONU were first encoded by $(17280, 14592)$-LDPC and mapped into 16QAM. The training sequence for the coherent DSP was added \cite{wang2023fast}. The pilots were inserted periodically into every 31 $(K - 1)$ payloads. After the pulse shaping using a square root-raised cosine (RRC) filter with a 0.1 roll-off factor, frequency shift, and subcarrier multiplexing, each ONU generated a 2-SCs$\times$8-Gbaud/SC DSCM signal.

The digital signal was converted to the analog signal by a 90-GSa/s digital-to-analog converter (DAC). The transmitted signal at each SC is 8 Gbaud. Each subcarrier contains 8640 16QAM payloads, 416-symbol training sequences, and 279 GCS-PAM4 pilots. Thus, the line rate of the secure coherent PON at US transmission is 256 Gbps (4-SCs$\times$8-Gbaud/SC$\times$4 bits/symbol/polarization$\times$2 polarizations), while the net rate is $\sim$200.08 Gbps ($8640/9335\times14592/17280\times$256 Gbps). The signal was amplified and modulated on the optical carriers @ $\sim$1549.85 nm for ONU 1 and $\sim$1550 nm for ONU 2 by a coherent driver modulator (CDM) with an output optical power of about $-12.08$ dBm. An Erbium-doped fiber amplifier (EDFA) was used as the booster amplifier with a launched optical power (LOP) of about $8$ dBm. EDFA should be replaced by a semiconductor optical amplifier for commercial deployment. Then the optical signals of two ONUs were coupled by a 50:50 optical coupler (OC). The optical spectrum of CDM output and the US signal at Point 1 with attenuation is shown in Fig. \ref{ex}(b). The optical signal was launched into a 20 km standard single-mode fiber (SSMF).

\begin{figure}[!t]
	\centering
	\includegraphics[width = \linewidth]{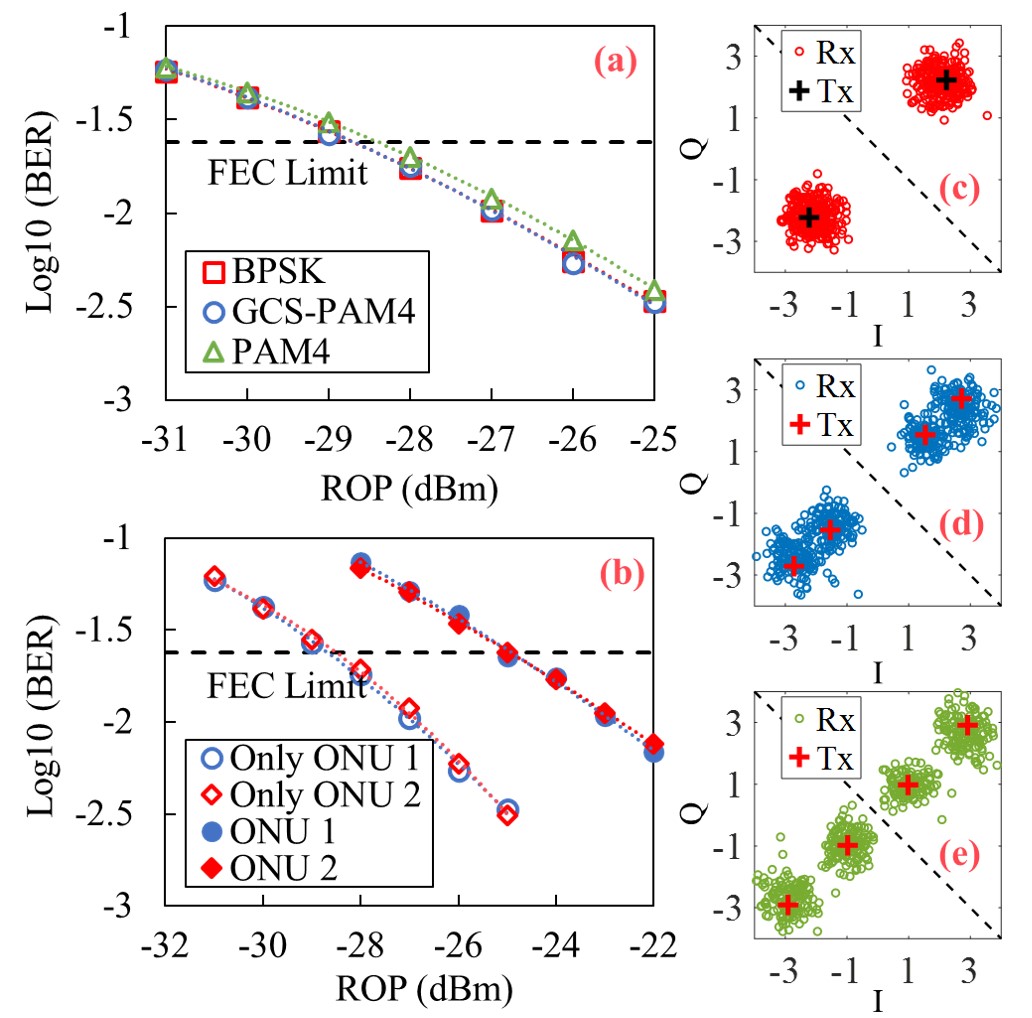}
	\caption{(a) Pre-FEC BER of the secure coherent PON using different pilots for CPR. (b) Pre-FEC BER of the ONU 1 and ONU 2 using the GCS-PAM4 pilot. Constellations of the (c) BPSK, (d) GCS-PAM4, and (e) PAM4 pilots.}
	\label{US}
\end{figure}

A variable optical attenuator (VOA) was used to adjust the received optical power (ROP) at the OLT. An external cavity laser (ECL) with a linewidth less than 100 kHz was used as the local oscillator (LO). The LO output power is $\sim$11.7 dBm and the wavelength is $\sim$1549.93 nm. The optical signal was mixed with the LO in an integrated coherent receiver (ICR) and converted to an analog signal. Then the signal was digitized by a 90-GSa/s analog-to-digital (ADC). The burst DSCM signal was recovered by the coherent DSP, including frame detection, frequency offset estimation (FOE), frequency shift, match filtering, timing recovery (TR), synchronization, multiple-input multiple-output (MIMO) equalizer, and the pilot-based CPR with a half-length $Q$ of 11. The 16QAM payload was de-mapped and LDPC decoded using the sum-product algorithm. Finally, the GCS-PAM4 pilot was recovered and the key bits were polar decoded using the CRC-aided successive cancellation list algorithm to realize the key distribution.

\subsection{AES algorithm at DS transmission}
The experimental setups of the secure coherent PON using 4-SCs$\times$8-Gbaud/SC DSCM with AES-encryption at the broadcast DS transmission are shown in Fig. \ref{ex}(c). The new keys from the US transmission for the AES algorithm were first updated. The bit sequence at OLT was encrypted by the AES-256 algorithm using the new keys of each ONU. After LDPC encoding, the bit sequence was mapped into 16QAM. Training sequence and pilot for CPR were also added for coherent DSP. Then a 4-SCs$\times$8-Gbaud/SC DSCM signal without guard interval was generated, including the power optimization, the pulse shaping using an RRC filter with a 0.1 roll-off factor, frequency shift, and subcarrier multiplexing.

Each subcarrier contains 8640 16QAM payloads, 480-symbol training sequences, and 279 pilots. Thus, the line rate of the secure coherent PON at DS transmission is 256 Gbps (4-SCs$\times$8-Gbaud/SC$\times$4 bits/symbol/polarization$\times$2 polarizations), while the net rate is $\sim$198.72 Gbps ($8640/9399\times14592/17280\times$256 Gbps). The DSCM signal was converted to an analog signal by the 90-GSa/s DAC and modulated at an optical carrier @ $\sim$1539.98 nm by a CDM with about $-12.93$ dBm output power. An EDFA was used as the boost amplifier at OLT for DS transmission. The LOP was optimized to $\sim$8 dBm by adjusting the gain of the EDFA. The optical spectrum of the signal at Point 1 and Point 2 is shown in Fig. \ref{ex}(d). Finally, the optical signal was launched into a 20 km SSMF.

A VOA was used to adjust the ROP at ONU. A tunable ECL with a linewidth of less than 100 kHz and $\sim$11.7 dBm output power was used as LO. The optical signal was converted to an analog signal by an ICR. The subcarriers of interest were selected by LO wavelength tuning and transferred into the baseband signal by coherent detection without using the optical filter. Then the analog signal was digitized by a 90-GSa/s ADC. The DSCM subcarriers of interest were recovered, including FOE, frequency shift, match filtering, TR, synchronization, MIMO equalizer, and pilot-based CPR. After the 16QAM payload was de-mapped and LDPC decoding, the AES decryption was implemented with the updated keys, which were generated by the ONU.

\begin{figure}[!t]
	\centering
	\includegraphics[width = \linewidth]{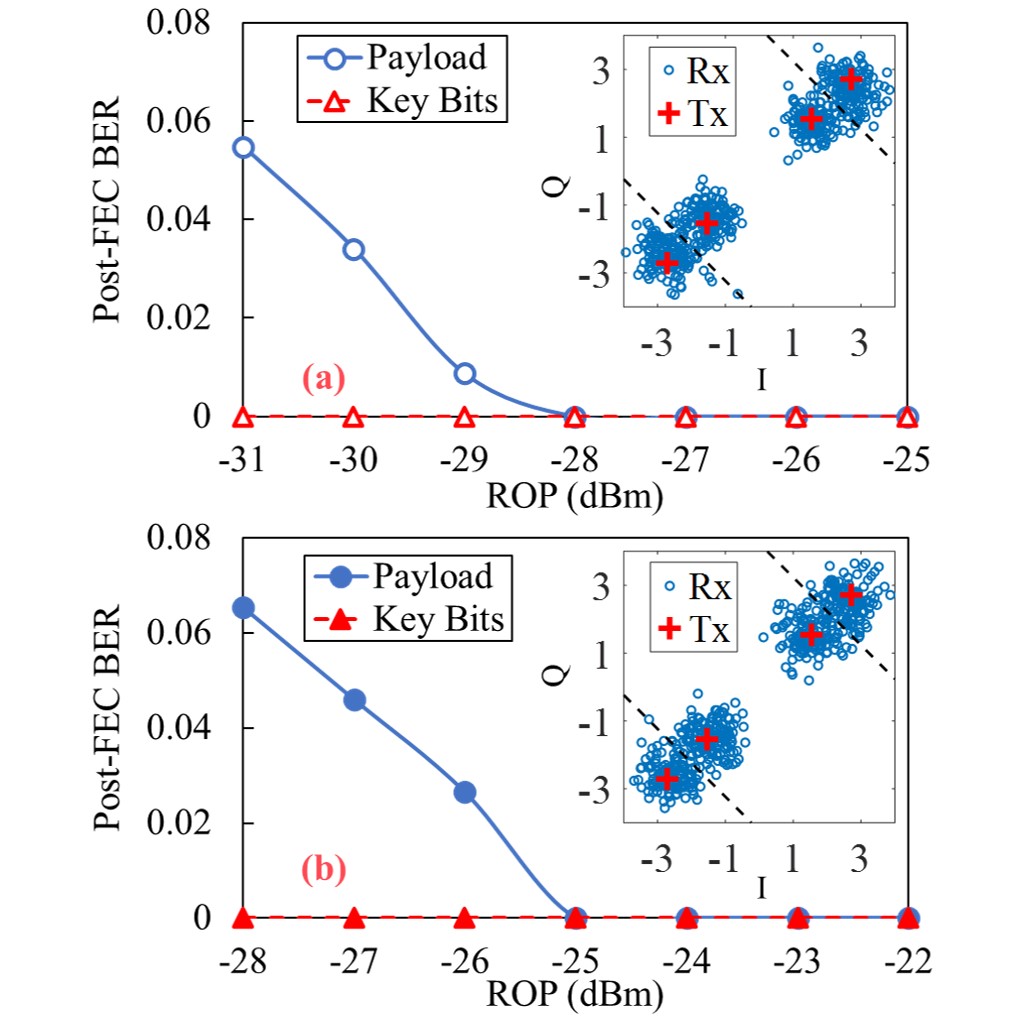}
	\caption{Post-FEC BER of the 16QAM payload and the key bit at (a) one-ONU and (b) two-ONU scenarios. Inset is the normalized GCS-PAM4 pilot at Tx and Rx.}
	\label{US_post}
\end{figure}

\section{Experimental results and discussions}
\label{RESULTS}
The pre-FEC BER of the secure coherent PON using different pilots for CPR at US transmission is shown in Fig. \ref{US}(a). Figs. \ref{US}(c-e) show the normalized constellations of the BPSK, GCS-PAM4, and PAM4 pilots at the transmitter (Tx) and receiver (Rx) with an ROP of $-28$ dBm, respectively. At the Tx, the power of PAM2, GCS-PAM4, and PAM4 pilots was normalized to be consistent with that of the payload. The dashed black lines are the decision thresholds of the first bits for CPR. Compared to the BPSK pilot for CPR, there is a $\sim$0.3 dB penalty on the receiver sensitivity at the 20\% SD-FEC limit when the PAM4 pilot is used for CPR. When the proposed GCS-PAM4 pilot is used for CPR, there is almost no penalty on the receiver sensitivity at the 20\% SD-FEC limit. Therefore, the proposed GCS-PAM4 pilot is effective for hardware-efficient CPR in the secure coherent PON. The pre-FEC BER of ONU 1 and ONU 2 using the GCS-PAM4 pilot is shown in Fig. \ref{US}(b). Scenarios that only ONU1 or ONU2 works alone and ONU1 and ONU2 work together were evaluated. The BER performance of the two ONUs is almost the same. When two ONUs work together, the signal bandwidth is doubled and the total power of the signal is 3 dB higher than that of the single ONU signal. The BER of the 16QAM payload is lower than the 20\% SD-FEC limit when the ROP is no less than $-28$ dBm and $-25$ dBm at one-ONU and two-ONU scenarios, respectively. It can be estimated that a 3 dB higher ROP is needed when the number of ONUs using FDMA is doubled and some penalties are caused by the out-of-the-band noise, while the system performance is slightly affected by burst detection when the ONUs use TDMA.

\begin{figure}[!t]
	\centering
	\includegraphics[width = \linewidth]{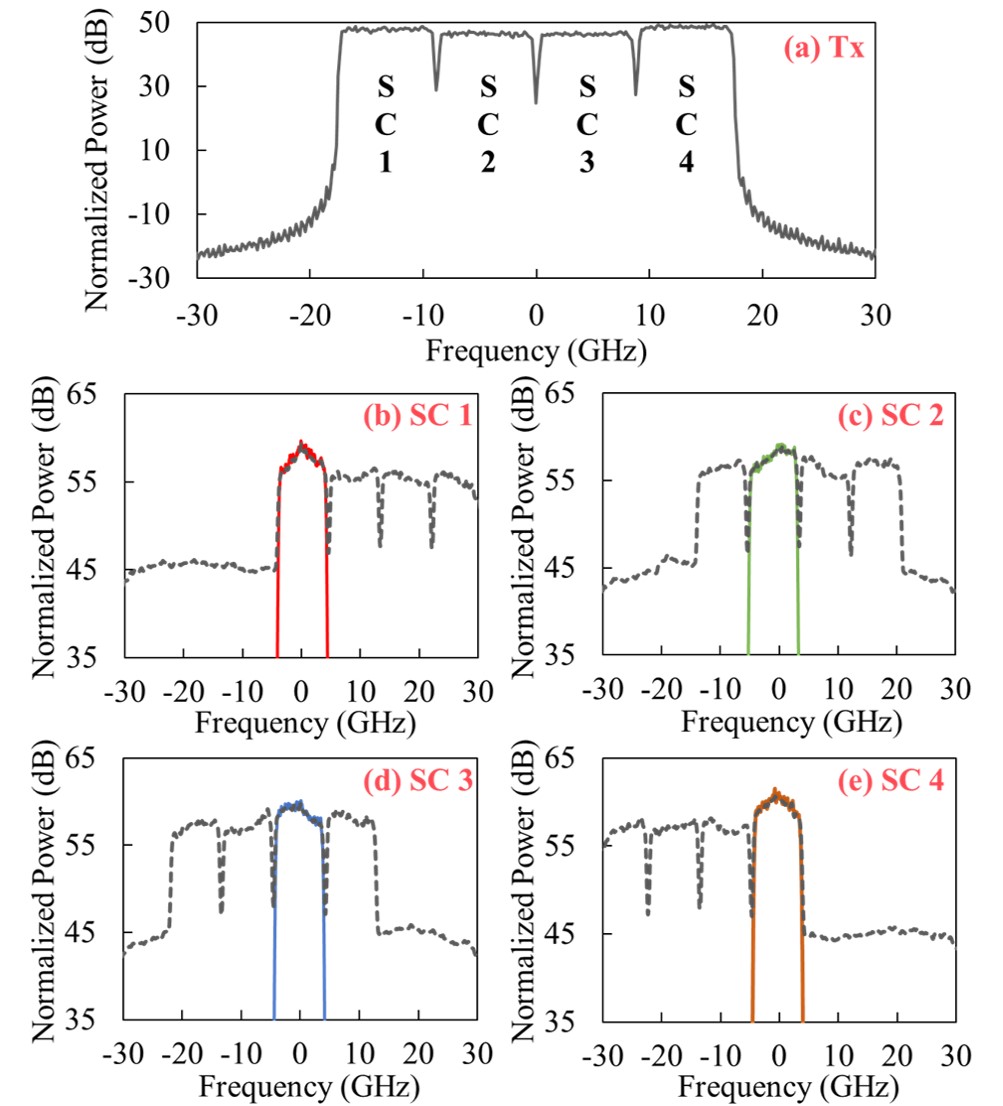}
	\caption{Electrical spectrum of the (a) transmitted DSCM signal, the received (b) SC 1, (c) SC 2, (d) SC 3, and (e) SC 4 in the coherent PON at DS transmission. The dashed lines represent the received signal and the solid lines represent the selected SC by wavelength tuning.}
	\label{DS_spectrum}
\end{figure}

The post-FEC BER of the 16QAM payload and the key bit for the AES algorithm at (a) one-ONU and (b) two-ONU scenarios are shown in Fig. \ref{US_post}. When the ROP is no less than $-28$ dBm and $-25$ dBm for the one-ONU and two-ONU scenarios, the 16QAM payload is error-free after the LDPC decoding. The power budget of the coherent PON using 4-SCs$\times$8-Gbaud/SC DSCM at US transmission is $\sim$33 dB (i.e., $+8$ dBm minus $-25$ dBm). The post-FEC BER of the 16QAM payload can match the pre-FEC BER in Fig. \ref{US}(b). Inset is the normalized GCS-PAM4 pilot at the Tx and Rx with ROPs of $-28$ dBm and $-25$ dBm, respectively. The dashed black lines represent the decision thresholds of the second bits for key distribution. Although the key bits with inadequate length become error-free at all the measured ROP after the polar code decoding, according to the simulation results, key bits should be error-free at an ROP of no less than around $-28$ dBm and $-25$ dBm, respectively. Thus, the GCS-PAM4 pilot is also effective for key distribution without additional overhead. The payload within one LDPC codeword is inserted by 279 pilots at each polarization to allow frequent key updates for every FEC codeword, which improves physical-layer security. Moreover, the proposed algorithm can be also applied to other encryption algorithms with the need for key distribution.

\begin{figure}[!t]
	\centering
	\includegraphics[width = \linewidth]{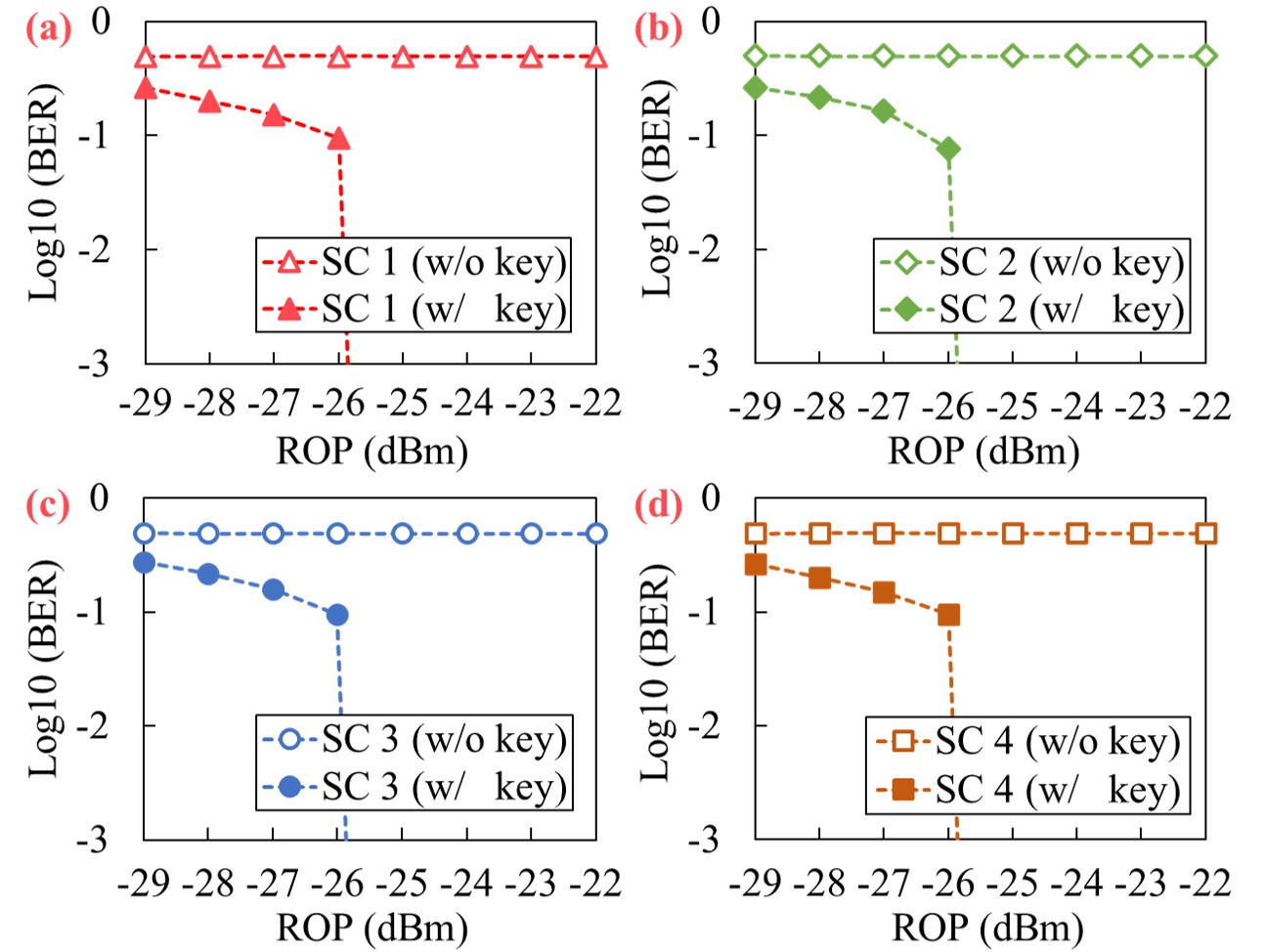}
	\caption{Post-FEC BER performance of (a) SC 1, (b) SC 2, (c) SC 3, and (d) SC 4 in the secure coherent PON using DSCM at DS transmission without or with key for AES decryption.}
	\label{DS_BER}
\end{figure}

The electrical spectrum of the (a) transmitted DSCM signal, the received (b) SC 1, (c) SC 2, (d) SC 3, and (e) SC 4 in the coherent PON at DS transmission is shown in Fig. \ref{DS_spectrum}. To maintain the similar BER performance of all SCs, more power was loaded to the SCs at a higher frequency, while those at a lower frequency were loaded with less power to compensate for the limited bandwidth of the transceiver. The dashed lines represent the received signal and the solid lines represent the selected SC. Each SC of the DSCM signal can be transferred to the baseband signal by tuning the wavelength of LO. Therefore, the coherent optical technologies and the DSCM for TFDMA allow the low-bandwidth transceivers at the ONUs. However, eavesdropping would still occur in the coherent PON using DSCM and it is not easily discovered by OLT. For example, illegal ONUs can eavesdrop on the broadcast DS signal of other ONUs at the same SCs or different SCs by wavelength tuning. Therefore, the physical-layer security enhancement is crucial for the future coherent PON using DSCM, which will also face severe security issues, especially eavesdropping at the broadcast DS transmission.

The post-FEC BER performance of each subcarrier in the secure coherent PON using DSCM at DS transmission without or with key is shown in Fig. \ref{DS_BER}. When the ROP is no higher than $-26$ dBm, the signal after LDPC decoding and the AES decryption with the correct key is still not error-free because the pre-FEC BER can not reach the 20\% SD-FEC limit. When the ROP is no less than $-25$ dBm, the signal of these four SCs after AES decryption with the correct key becomes error-free. Since the power of the subcarriers is adjusted to balance the SNR, the BER performance of these four subcarriers is similar. The power budget of the coherent PON at DS transmission is $\sim33$ dB (i.e., $+8$ dBm minus $-25$ dBm). However, despite the increase in the ROP, the post-FEC BER of the four SCs is still $\sim$0.5 when the correct key of the AES decryption is unavailable for illegal ONUs. Therefore, the physical-layer security of the coherent PON is enhanced by the proposed scheme based on the AES algorithm with frequent key updates using the GCS-PAM4 pilot. It is worth mentioning that the eventual timing processing is asymmetry in the DS and US transmission due to pilot mapping for CPR and key generation/data encryption.

\section{Conclusion}
\label{CONCLUSION}
In this paper, we propose the AES algorithm and GCS-PAM4 pilot-based key distribution for secure coherent PON. At the unicast US transmission, the first bit of the proposed GCS-PAM4 pilot is used for the hardware-efficient CPR and the second bit is encoded by the polar code and utilized for error-free key distribution without additional overhead. Frequent key updates are permitted for every codeword to improve the security of coherent PON at the broadcast DS transmission. Experimental results of the 200-Gbps secure coherent PON using DSCM demonstrated that the polar code and GCS-PAM4 pilot enable the error-free key distribution without additional overhead at US transmission, while the AES-256 algorithm has prevented eavesdropping at DS transmission. Furthermore, the proposed GCS-PAM4 pilot exhibits comparable CPR performance to the BPSK pilot with the best CPR performance. In conclusion, the proposed physical-layer security enhancement scheme shows potential in the application to future coherent PON.


\bibliographystyle{IEEEtran}
\bibliography{sample}

\vfill

\end{document}